\documentclass[12pt,nofootinbib]{revtex4}
\usepackage{amsmath}
\usepackage{amsfonts}
\usepackage{subfigure}
\usepackage{graphicx}
\usepackage{psfrag}
\usepackage{tikz}

\newcommand{\eref}[1]{Eq.~(\ref{#1})}

\newcommand{\nn}{\nonumber}

\newcommand{\fref}[1]{Fig~\ref{#1}}

\def\be{\begin{equation}}
\def\ee{\end{equation}}

\def\bea{\begin{eqnarray}}
\def\eea{\end{eqnarray}}

\ifx\du\undefined
  \newlength{\du}
\fi
\begin{document}
\title{Supercurrent: Vector Hair for an AdS Black Hole}

\author{Pallab Basu}
 \email{pallab@phas.ubc.ca}
 
 \author{Anindya Mukherjee}
 \email{anindya@phas.ubc.ca}

\author{Hsien-Hang Shieh}
\email{shieh@phas.ubc.ca}
 
 \affiliation{Department of Physics and Astronomy, 
 University of British Columbia,
 6224 Agricultural Road,
 Vancouver, B.C. V6T 1Z1,
 Canada}

\begin{abstract}

 In \cite{Hartnoll:2008vx} a holographic black hole solution is discussed which exhibits a superconductor like transition. In the superconducting phase the black holes show infinite DC conductivity. This gives rise to the possibility of deforming the solutions by turning on a time independent current (supercurrent), without any electric field. This type of deformation does not exist for normal (non-superconducting) black holes, due to the no-hair theorems. In this paper we have studied such a supercurrent solution and the associated phase diagram. Interestingly, we have found a ``special point'' (critical point) in the phase diagram where the second order superconducting phase transition becomes first order. Supercurrent in superconducting materials is a well studied phenomenon in condensed matter systems. We have found some qualitative agreement with known results.  
\end{abstract} 

\maketitle

\section{Introduction}

The AdS/CFT correspondence \cite{Maldacena:1997re} has proved to be one of the most fruitful ideas in string theory. It has provided important insights into the nature of strongly coupled gauge theories. Various field theoretic phenomena like confinement/deconfinement transition \cite{Witten:1998zw}, chiral symmetry breaking \cite{Sakai:2004cn}, transport properties \cite{Son:2007vk}, etc have been understood from a string theory (gravity) view point. In recent times, there have been various works which aim at constructing string theory (gravitational) duals to condensed matter systems \cite{Herzog:2007ij,Hartnoll:2007ih,Balasubramanian:2008dm,Adams:2008wt,Maldacena:2008wh,Kachru:2008yh,Herzog:2008wg,Hartnoll:2008hs,Davis:2008nv}.

A superconductor (or BEC in general) is one of the most studied systems in condensed matter physics \cite{Tinkham} and is also equally important in the study of the phase diagram of QCD \cite{Rajagopal:2000wf}. However there are certain aspects of superconductivity, e.g. high temperature superconductivity etc, that are still not completely understood. As superconductivity is a field theoretical phenomenon, it is interesting to ask if gauge gravity duality can be used to provide some insights into superconductivity. It turns out that there exists a gravitational system which closely mimics the behaviour of a superconductor. We will briefly describe the basic set up. Recently it has been shown by Gubser that in the $AdS_4$ background one can have condenstation of a charged scalar field \cite{Gubser:2008px}\footnote{Possible ways to have scalar hair in asymptotically ﬂat EYMH systems (also related mechanism in non-abelian gauge gravity systems) has been discussed by authors (see \cite{Bartnik:1988am,Bizon:1990sr,Mavromatos:1995fc,Volkov:1998cc} and references therein).}. It is shown that there exist solutions that allow for a condensing scalar to be coupled to the black hole if the charge on the black hole is large enough. The scalar couples to a $U(1)$ gauge field under which the black hole is charged, and its condensation breaks
the gauge symmetry spontaneously, giving a mass to the gauge field. The exact backreacted gravity solution with the condensation of scalar field is difficult to find. In \cite{Hartnoll:2008vx} the gauge fields and coupled scalar part of the Lagrangian has been solved and studied numerically, neglecting the gravity backreaction. It has also been shown that system undergoes a second order normal conductor/DC superconductor transition after the scalar condenses. Various related works discuss other aspects of superconductivity including partial discussions of Meissner effect and the non-abelian case \cite{Gubser:2008zu,Gubser:2008wz,Roberts:2008ns,Gubser:2008wv,Wen:2008pb,Albash:2008eh,Maeda:2008ir}.

Here we carry these investigations further by presenting a DC supercurrent type solution. Infinite DC conductivity in the dual field theory means that there are states in the field theory with time independent non-zero DC current (but without any external voltage). Such a phenomenon is well known in condensed matter systems. It is known from various experiments that the supercurrent can sustain itself for several years in a superconducting coil. In terms of AdS/CFT the above mentioned supercurrent states will correspond to a deformation of superconducting black holes by the spatial component of the gauge fields with a non-trivial radial dependence. We have numerically constructed such a solution. The solution may be thought as a vector hair to a superconducting AdS black hole and may be interpreted as a bound state of soliton and a black hole. As one would expect, a type of no-hair theorem \cite{Bekenstein:1996pn} prevents any such non-trivial solution from occurring in the case of a normal (non-superconducting) AdS black hole. 

\begin{figure}
\begin{center}
\includegraphics[scale=0.4]{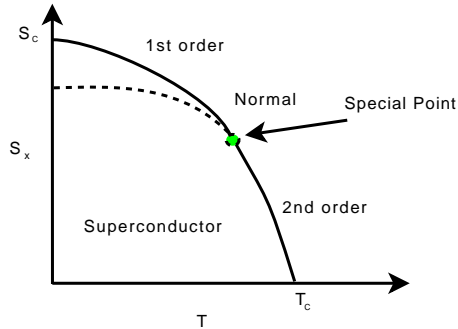}
\end{center} 
\caption{Phase diagram in $S_x,T$ plane showing critical point, first order and second order transition. For $T<T_{sp}$ the phase transition is first order. The dotted line is the extension of second order transition line.}
\label{fig:phase}
\end{figure}

We also have studied the interesting and novel phase diagram of such a system. It is shown that the critical temperature of the superconducting transition decreases with the introduction of a chemical potential for supercurrent and most interestingly at some point the order of phase transition changes from second order to first order. Thereby we have shown the existence of a ``special point''\footnote{This is actually a critical point. However we refrain from using the word ``critical'' to avoid any possible confusion with $T_c,S_c$ etc.} in the phase diagram, where the line of second order transitions ends and the first order transition begins (\fref{fig:phase}).
It has also been shown that the superconducting phase transits back to the normal phase at a certain value of the supercurrent (critical current). The associated transition may be second order or first order depending on temperature.

In this paper we start by discussing the equations and general setup (section \ref{EABC}), then we discuss the superconducting black hole solution studied in \cite{Hartnoll:2008vx} and also introduce our supercurrent solution (section \ref{SOL}). In section \ref{RESULTS} we discuss the various phase diagrams associated with our model. We also comment on some possible connection with superfluid phase diagram (section \ref{PHYREL}) and the issue of gravity backreaction (section \ref{BCKR}). In the concluding section we discuss some future directions and open questions.

{\bf Note added:} After this work was completed we learned of a work by C.P.Herzog, P.K.Kovtun and D.T.Son titled ``Holographic model of Superfluidity'' which has some overlap with our results. We thank Kovtun for informing us about this work.

\section{Equations and Accounting of the Boundary Conditions } 
\label{EABC}

 In this section we will describe the setup on the gravity side which gives a superconducting system in the boundary theory.  Following \cite{Hartnoll:2008vx}, we consider the planar limit of the four dimensional AdS black hole:
\begin{equation}
 ds^2 = -f(r) dt^2 + \frac{dr^2}{f(r)} + r^2 (dx^2 + dy^2)
\end{equation}
where
\begin{equation}
f(r) =\frac{r^2}{L ^2}-\frac{M}{r}
\end{equation}
and  $L$ is the radius of the anti-de Sitter space and the temperature of the black hole (and also the boundary field theory) is given by
\begin{equation}
 T =\frac{3M^{1/3}}{4\pi L^{4/3}}
\end{equation}
 In this note we will adopt the convention that $M =L =1 $.  At the phenomenological level, superconductivity is usually modeled by a Landau-Ginzburg Lagrangian where a complex scalar field develops a condensate in a superconducting phase. In order to have a scalar condensate in the boundary theory, the authors of \cite{Hartnoll:2008vx} introduce a $U(1) $ gauge field and a conformally coupled charged complex scalar field $\psi $ in the black hole background. The Lagrangian of the system is:
\begin{equation}
 L =\int dx^4 \sqrt{-g} ( -\frac{1}{4}F^{a b}F_{ab}-V(|\psi |) -|\partial\psi -iA\psi |^2),
\end{equation}
where the potential is given by
\begin{equation}
 V(|\psi|) =-2 \frac{|\psi|^2}{L^ 2}
\end{equation}
which corresponds to the conformal mass $m_{conf}^2 = -2/L^2$. As argued in \cite{Hartnoll:2008vx}, the mass term is negative but above the Breitenlohner-Freedman (BF) bound \cite{Breitenlohner:1982jf} and thus does not cause any instability in the theory. As we will see below the presence of the vector potential effectively modifies the mass term of the scalar field as we move along the radial direction $r$ and allows for the possibility of developing hairs for the black hole in parts of the parameter space. Noticed that in our model there's no explicit specification of the Landau-Ginzburg potential for the complex scalar field. The development of a condensate relies on more subtle mechanisms for violations of the no hair theorem. 

Here we consider the possibility of a DC supercurrent in this setup. For this purpose, we will have to turn on both a time component $A_t $ and  a spatial component $A_x$ for the vector potential. We are interested in static solutions and will also assume all the fields are homogeneous in the field theory directions with only radial dependence. It is more convenient to analyze the system by making a coordinate transformation $z = 1/r $. The metric becomes:
\begin{equation}
 ds^2 = -f(z) dt^2 + \frac{dz^2}{z^4 f(z)} +  \frac{1} {z^2}  (dx^2 + dy^2)
\end{equation}
with
\begin{equation}
f(z) =\frac{1}{z ^2}-z.
\end{equation}
The horizon is now at $z=1$, while the conformal boundary lives at $z=0$. Like \cite{Hartnoll:2008vx} we will also neglect the gravity back reaction of gauge and scalar fields. How this limit can be taken consistently is discussed in \cite{ Hartnoll:2008vx,Gubser:2008zu}. The equations of motion for the fields in this coordinate system are:
\begin{eqnarray}\label{maineq}
  \psi ''&+&  \frac{ f'}{f} \psi'+ \frac{1}{ z^4}   \left( \frac{  A_t ^2  }{ f^2  } -\frac{z^2  A_x ^2 }{  f }  + \frac{2}{L^2  f} \right)\psi =0 \\
\nn  A_t ''&-&  2\frac{ \psi^2} { f  z^4}   A_t =0 \\
\nn  A_x ''&+&  \left ( \frac{  2  }{ z  } + \frac{  f'}{  f }  \right) A_x' - 2\psi ^2    \frac{  A_x  }{ z^4 f } =0
\end{eqnarray}
To require regularity at the horizon we will have to set $A_t =0  $ at $z=1 $. Since we have a set of coupled equations,
this will in turn give the constraints at the horizon
\begin{eqnarray}
\label{regular}
  z\psi '&=&  \frac{ 2}{3} \psi-\frac{1}{3}z^2 A_{x}^2 \psi^2 \\
\nn  A_x '&=&       -\frac{ 2  }{ 3}(\frac{  \psi }{ z  })^2    A_x \\
\nn A_t&=&0
\end{eqnarray}
where $z=1$.
Examining the behaviour of the fields near the boundary, we find
\begin{eqnarray}
  \psi &\sim &  \Psi_1 z+ \Psi_2  z^ 2  + ...\\
\nn  A_t &\sim &  \mu - \rho z + ...\\
\nn  A_x &\sim &  S_x  + J_x z + ...
\label{asymp}
\end{eqnarray}
The constant coefficients above can be related to physical quantities in the boundary  field theory using the usual dictionary in gauge/gravity correspondence.
$\mu $, $\rho$ are the chemical potential and the density of the charge carrier in the dual  field theory, respectively. $J_x$ corresponds to the current, while $S_x$ gives the dual current source. $\Psi_{1,2} $ are both coefficients multiplying normalizable modes of the scalar  field equation. They are the expectation values of operators in the  field theory.
\begin{equation}
  \Psi_i \sim  <\mathcal O_i >
\end{equation}
In this paper, we will mainly study the $\Psi_1=0$ case and also briefly discuss $\Psi_2=0$ case. It turns out both exhibit similar behavior when it comes to DC superconductivity.

We want to parametrize our solutions in terms of dimensionless quantities. From the analysis in the appendix we see that $T $, $\mu $, $\Psi_1 $, $S_x  $ have dimension one, while $\rho$, $\Psi_2 $, $J_x $ have dimension two. The dimensionless combinations are $( \frac{T}{\mu }, \frac{S_x}{\mu },\frac{J_x}{\mu^2}, \frac{\sqrt{<\mathcal O_2 >}}{\mu },\frac{<\mathcal O_1>}{\mu})$. With the regularity conditions (\eref{regular}) and $\Psi_1 =0$ (or $\Psi_2=0$) , we are left with a two parameter family of solutions; which we characterize by two dimensionless quantities $\frac{T}{\mu},\frac{S_x}{\mu}$. 
Since the effect of temperature is governed by $\frac{T}{\mu}$, we can in practice keep the temperature fixed and achive the same effect by changing $1/\mu$. 
\section{Nature of the Solution} \label{SOL}
\subsection{Superconducting black hole}
\begin{figure}
\begin{center}
 \includegraphics[scale=0.5]{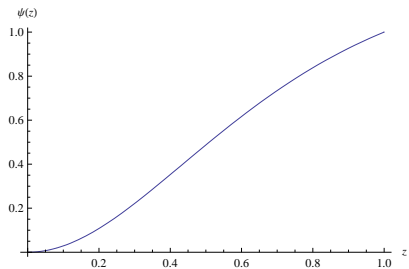}
\end{center}
\caption{Zero mode of $\psi$ at $\mu=\mu_c$ with a normalization $\psi=1$ at the horizon.}
\label{fig:B0Inst}
\end{figure}

Here we discuss the space-time profile of the superconducting black hole solution found in \cite{Hartnoll:2008vx} and put $A_x=0$ in the \eref{maineq}. At a small value of $\mu$ the only solution to the set of equations \ref{maineq} is given by,
\bea
\label{ordinary}
A_t &=& \mu (1-z) \\
\nn \psi &=& 0 
\eea
The effective mass of the field $\psi$ in this background is given by
\bea
m^2_{eff}=-2-\frac{A_t^2}{f(z)}
\eea
Hence, as we increase $\mu$  the effective mass becomes less than BF bound in a sufficiently large region of space and consequently a zero mode develops for the field $\psi$ (\fref{fig:B0Inst}) at $1/\mu=1/\mu_c \approx 0.146$. As $\mu$ is increased further the field $\psi$ condenses and a new branch of solution shows up which has non-zero value of $\psi$. (As discussed in \eref{asymp}, there are two possible boundary condition for $\psi$. Here we concentrate on $\Psi_1=0$ case, $\Psi_2=0$ case is similar.) It has been argued in  \cite{Hartnoll:2008vx} that this new solution has a lower free energy than the regular black hole solution with $\psi=0$ and there is a second order phase transition associated with this phenomenon. We present the phase diagram in the next chapter which may be thought as the $A_x=0$ case in our context\footnote{As we increase $\mu$ further there is a possibility of multi-nodal (in radial direction) solution, but as in \cite{Hartnoll:2008vx} we will not consider such solutions. These solutions probably are thermodynamically unfavourable.}. Here we will present the general nature of the solution. 

\begin{figure}
\begin{center}
\subfigure [Plot of $\psi(z)$. ] {\label{fig:psiplot}\includegraphics[scale=0.5]{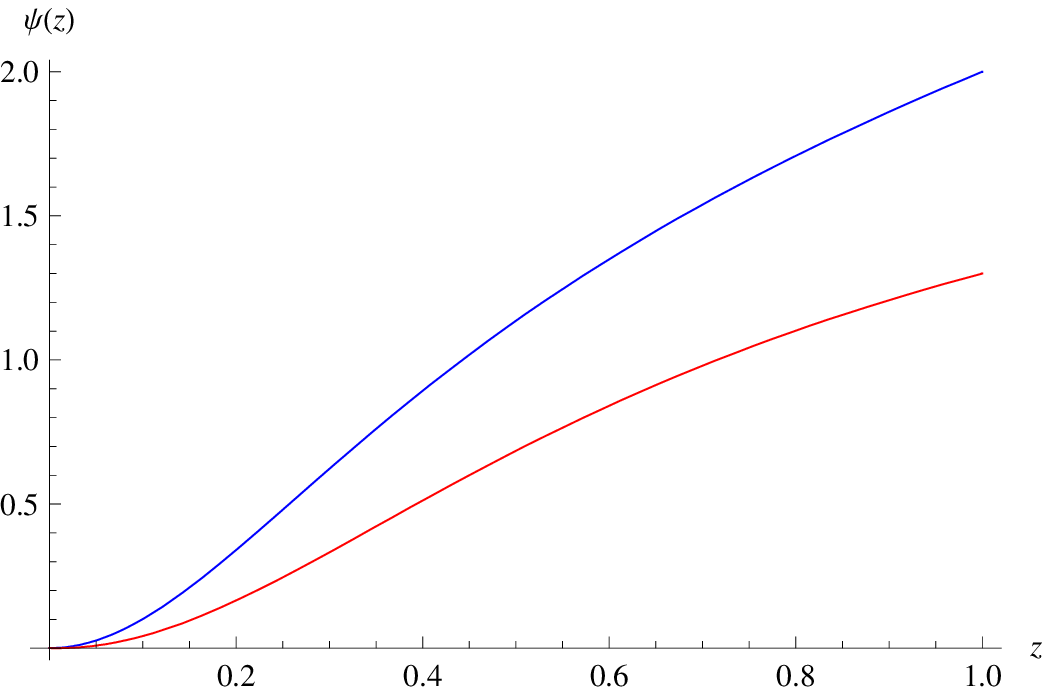}}\hspace{0.5cm}
\subfigure [Plot of $A_t(z)$, the curved ones. The straight lines are the plots of $A_t(z)$ for $\psi=0$ case with the same value of $1/\mu$.]{\label{fig:Atplot} \includegraphics[scale=0.5]{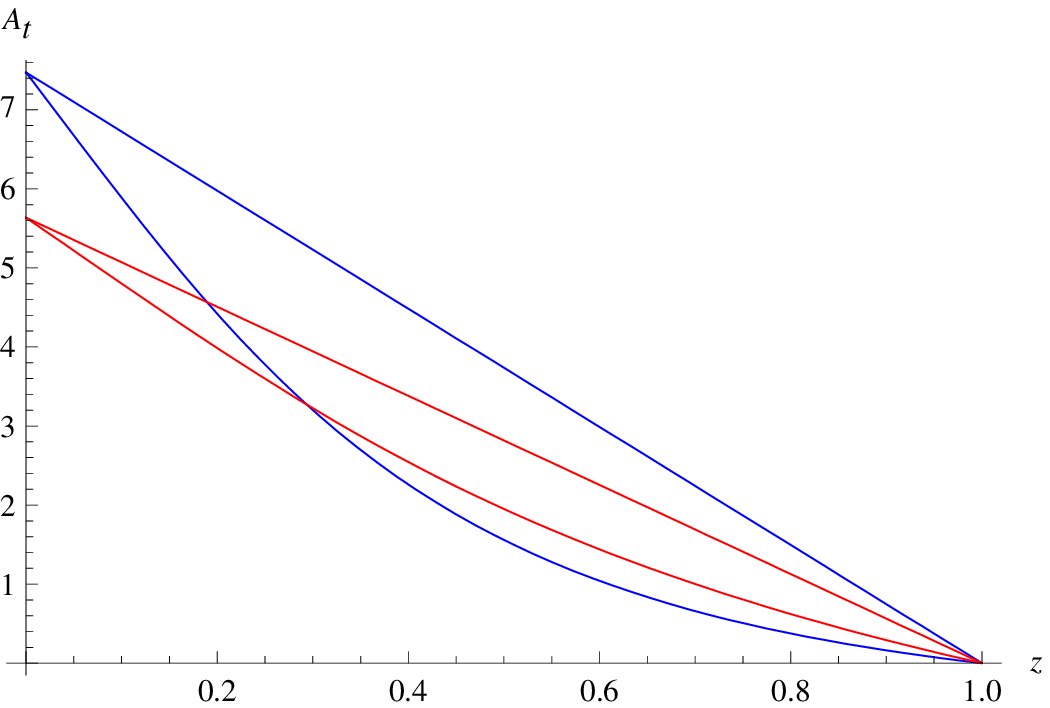}}
\end{center}
\caption{Plots of $\psi$ and $A_t$ at $1/mu \approx 0.105, 0.079$. $\mu$ is increasing from below. }
\end{figure}

We plot the solution for some generic values of $1/\mu \approx 0.105,0.079$ ( \fref{fig:psiplot}, \fref{fig:Atplot} ).  As one can see, $\Psi_2$ increases as we increase the value of $\mu$\footnote{Another interesting property is that as we increase $\mu$, $\rho=A'_t(z)$ at the boundary $z=0$ increases, however the charge of the black hole $\rho=A'_t(z_0)$ decreases. Although the the total charge of the configuration is increased by increasing $\mu$, the condensation becomes more dense and carries most of the charge.}.

The conductivity of this system can be calculated by looking at the frequency ($\omega$) dependent fluctuation equation of $A_x$ in this back ground. It has been shown in \cite{Hartnoll:2008vx} that imaginary part of conductivity has a pole at $\omega=0$. Consequently the real part of the conductivity will have a delta function at $\omega=0$. This example of an infinite DC conductivity is an example of superconductivity. In contrast to that an ordinary black hole (with $\psi=0$) has a finite DC conductivity. 

\subsection{Supercurrent solution}
The conductivity is given by (from \eref{asymp})
\bea
\sigma=\frac{J_x}{\dot{S_x}}
\eea
An infinite value of conductivity (as discussed at the end of the previous chapter) implies $J_x$ may be non-zero even if the $\dot{S_x}=0$. This is natural to expect in a superconductor, that current may flow without any applied voltage. Hence one may expect that one can deform the superconducting black hole solutions by turning on non-zero $J_x$ (consequently a non-zero $A_x$). Here we have constructed such a solution numerically by solving the coupled \eref{maineq}. As we are solving the coupled equation our solution is valid for any value of $A_x$ within the approximation scheme of neglecting gravity backreaction. The solution is characterized by two chemical potentials $S_x, \mu$ which are the boundary values of the fields $A_x$ and $A_t$ \eref{asymp}. To find this solution we start with fields satisfying appropriate boundary conditions near the black hole horizon \eref{regular} and integrate up to the boundary $z=0$. Just like the case of superconducting black hole solution we can put either $\Psi_1$ or $\Psi_2$ to zero. Here we present the solution with $\Psi_1=0$, the case $\Psi_2=0$ is similar. 

It should be noted that such a solution does not exist in an ordinary black hole background. From \eref{maineq}  we have for $\psi=0$ case 
\begin{eqnarray}
& & A_x '' +( \frac{  2  }{ z  } + \frac{  f'}{  f }  ) A_x  = 0 \\
&\Rightarrow& z^2 f A_x'' + (z^2 f)' A_x=0 \\
&\Rightarrow&  z^2 f (z^2 f A_x')'=0 \\
&\Rightarrow&  \frac{d^2}{dy^2} A_x=0, \quad dy=\frac{dz}{z^2 f}\\
&\Rightarrow& A_x = c_1+c_2 y
\end{eqnarray}

Near the horizon $z^2 f(z)=-3(1-z) +\dots$ Hence, 

\begin{eqnarray}
\nn y &\sim& \log(1-z) \\
\Rightarrow F_{zx} &\sim& \partial_z {A_x} \sim c_2 \frac{1}{1-z}
\end{eqnarray}
Energy density near the black hole horizon has a contribution from $g^{zz}g^{xx}F_{zx}^2$ term and consequently diverges as $1/(1-z)$ near the black hole horizon. Hence the only possible finite energy solutions are a constant ($c_2=0$,$z$ independent) $A_x$ solution with other fields given by \eref{ordinary}. As discussed  $J_x=0$ for such solutions. Free energy competition between such a normal solution and supercurrent solution gives rise to an intricate phase diagram. Actually the order of superconducting phase transition changes from second order to first order as we tune the boundary value of $A_x$. It should also be noted that the introduction of field $A_x$ changes the effective mass of the field $\psi$,
\bea
m^2_{eff}=-2-\frac{A_t^2}{f(z)}+z^2 A_x ^2
\eea
This implies that the introduction of too much $A_x$ may destroy superconductivity and suggests the possibility of a cortical value of $S_x$, beyond which there is no superconductivity. These issues related to phase transition has been discussed in section \ref{RESULTS}. Here we will show how the solution looks like. 

\begin{figure}
 \begin{center}
 \subfigure [Plot of $\psi$] {\label{fig:plot2B} \includegraphics[scale=0.5]{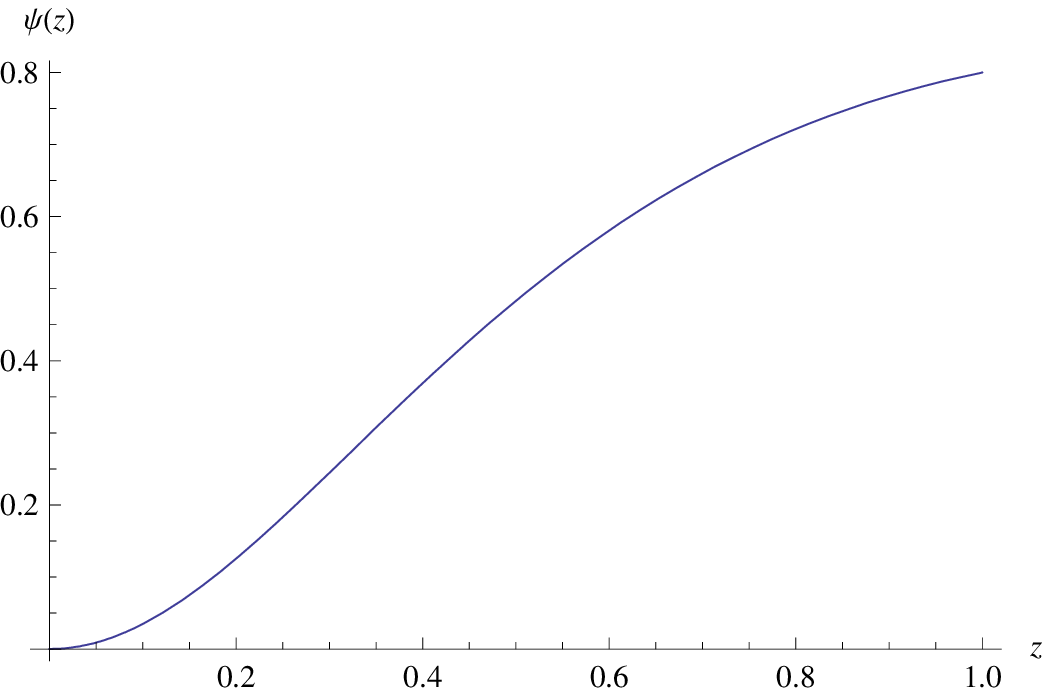}}
 \subfigure [Plot of $A_x$] {\label{fig:plotB2B} \includegraphics[scale=0.5]{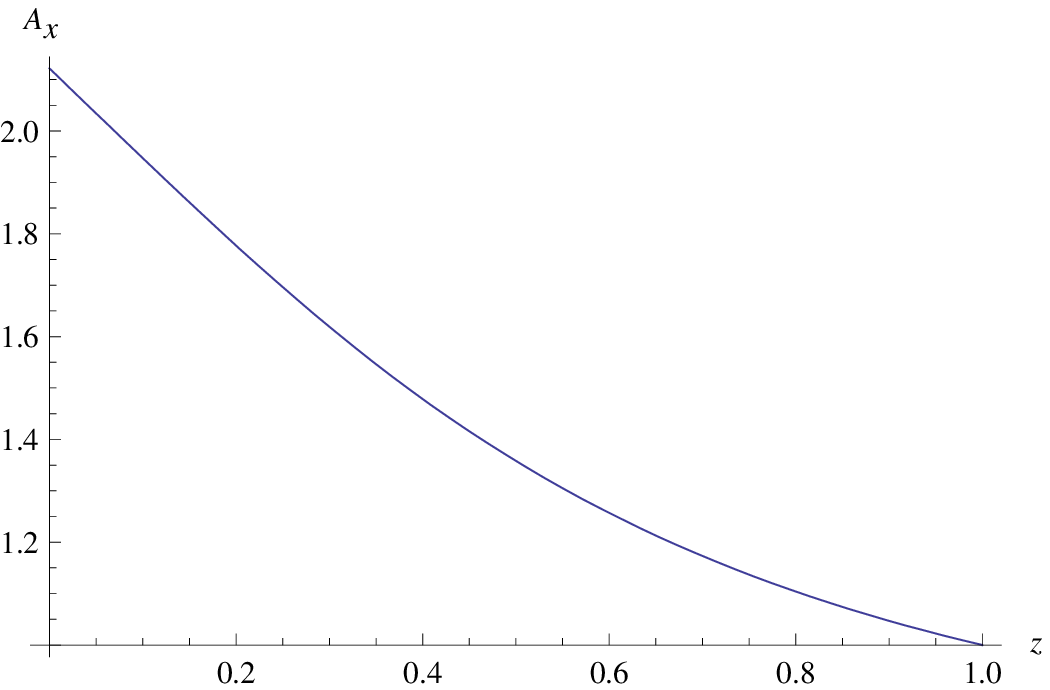}}
\end{center}
\caption{Nature of solution for $\frac{1}{\mu} \approx 0.174$ and $\frac{S_x}{\mu} \approx 0.369$}
\end{figure}

\begin{figure}
 \begin{center}
\subfigure [Plot of $\psi$] {\label{fig:plot1B}  \includegraphics[scale=0.5]{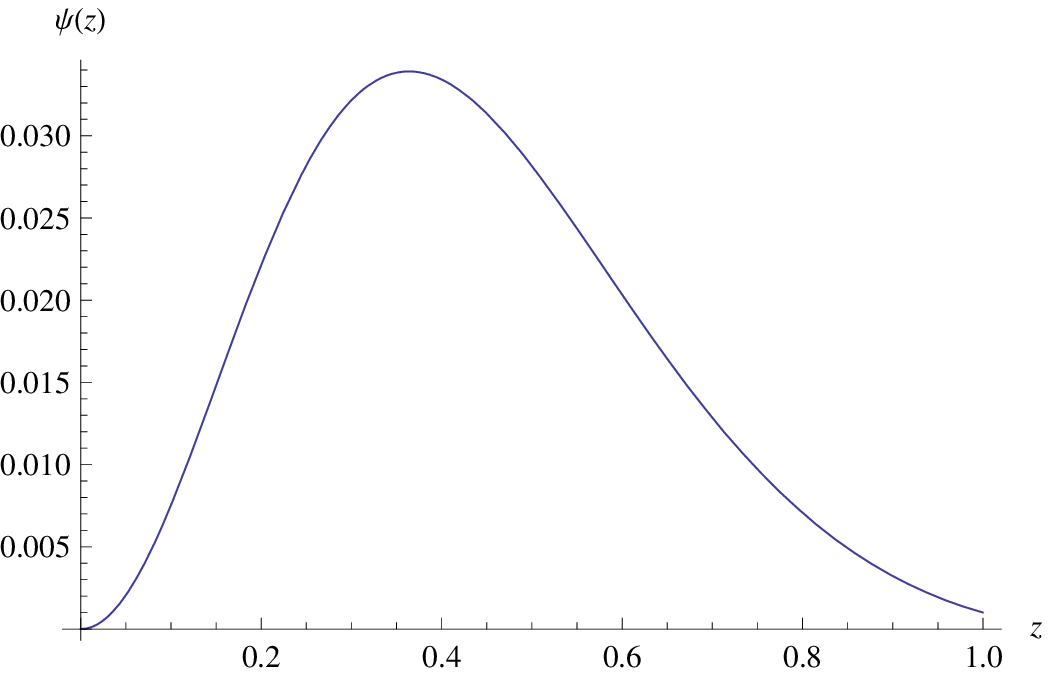} \hspace{0.5cm}}
\subfigure [Plot of $A_x$] {\label{fig:plotB1B} \includegraphics[scale=0.5]{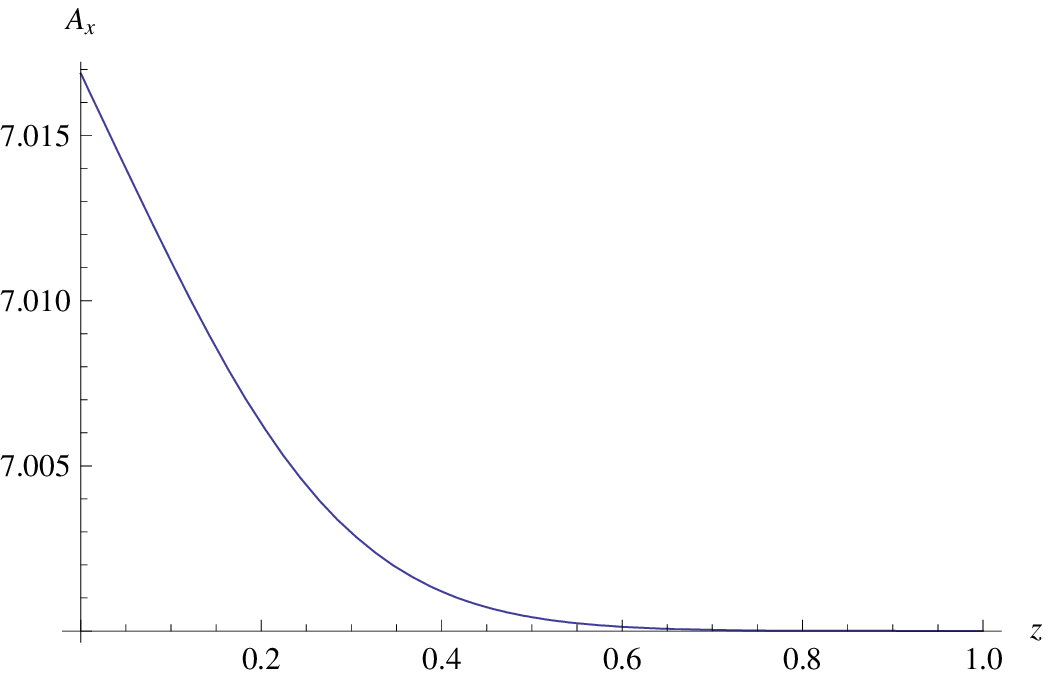}}
\end{center}
\caption{Nature of solution for $\frac{1}{\mu} \approx 0.087$ and $\frac{S_x}{\mu} \approx 0.609$ }
\end{figure}

From \eref{regular}, one may argue that the slope of the field $\psi$ at the horizon changes sign as one turns on $A_x$. This may be seen from the solutions with generic conditions $1/\mu \approx 0.174,S_x/\mu \approx 0.369$ (Fig \ref{fig:plot2B},\ref{fig:plotB2B}) and $1/\mu \approx 0.087,S_x/\mu \approx 0.609$ (Fig \ref{fig:plot1B},\ref{fig:plotB1B}). As we turn on more $S_x$ the value of $\psi$ at the horizon decreases. \fref{fig:plot2B} shows one configuration which is near the phase boundary. Turning on $S_x$ further will eventually destroy the condensate.

\section{Results}
\label{RESULTS}

\subsection{$\Psi_1 = 0$}
In this section we discuss the phase diagram associated with our solution. Let us for the moment set $\Psi_1 = 0$. This corresponds to choosing a boundary condition such that $\psi(z) \sim z^2$ at the boundary ($z \rightarrow 0$).

\begin{figure}
  \begin{center}
  \includegraphics[width=8cm]{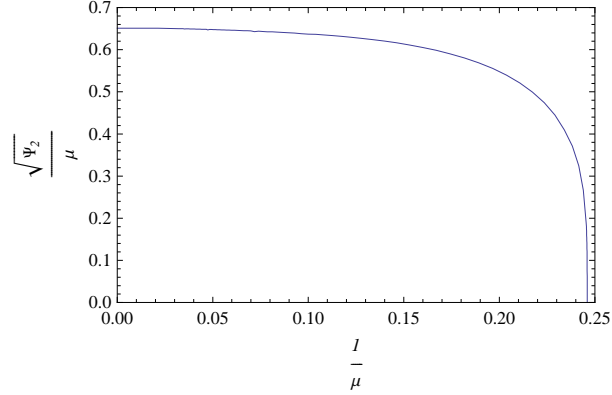}
  \end{center}
\caption{Plot of $\Psi_2$ as a function of $1/\mu$, for $A_x = 0$.}
\label{fig:psi2curve}
\end{figure}

 We first look at the case when $A_x = 0$. Solving for the condensate strength as a function of temperature or, equivalently, $1/\mu$ we get the curve shown in Fig \ref{fig:psi2curve}. For small values of $1/\mu$ the condensate strength reaches a saturation value. Near the point $\Psi_2 = 0$ the curve has the dependence $\sqrt{\Psi_2}/\mu \sim (1/\mu - 1/\mu_c)^{1/2}$, as expected. This corresponds to a second order phase transition. If the parameter $1/\mu$ is increased further the condensate ceases to exist, i.e., $\psi(z) = 0$ beyond this point. The critical value $1/\mu_c = 0.246$. For a fixed boundary value of $\frac{1}{\mu}$ Solutions with $\Psi_2 \neq 0$ always has less free energy than the normal black hole solution given in \eref{ordinary}.

\begin{figure}[h!]
 \begin{center}
   \subfigure[$1/\mu = 0.146$]{\label{fig:phasecurve1} \includegraphics[width=5.5cm]{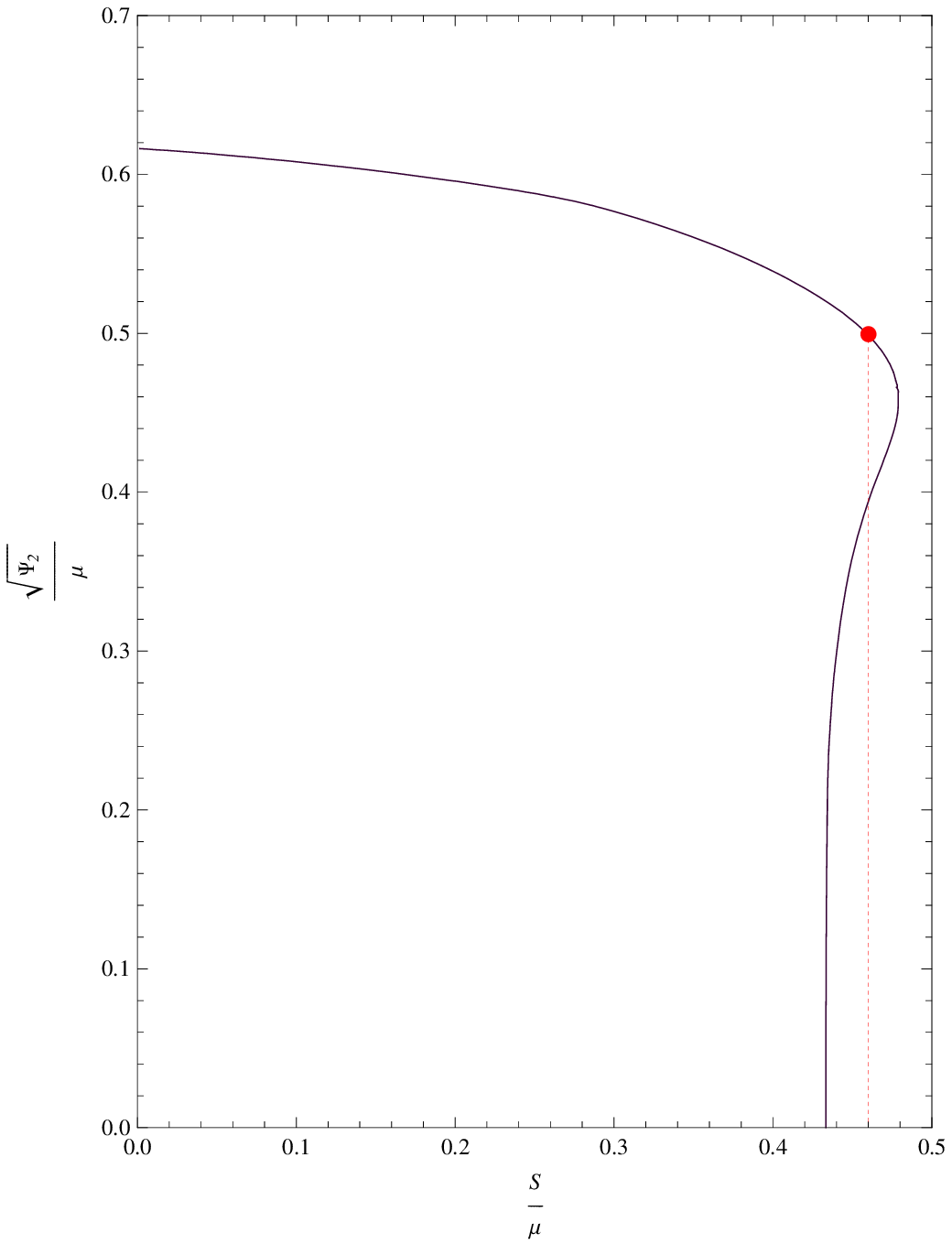}} \hspace{0.1cm}
   \subfigure[$1/\mu = 0.217$]{\label{fig:phasecurve2} \includegraphics[width=5.5cm]{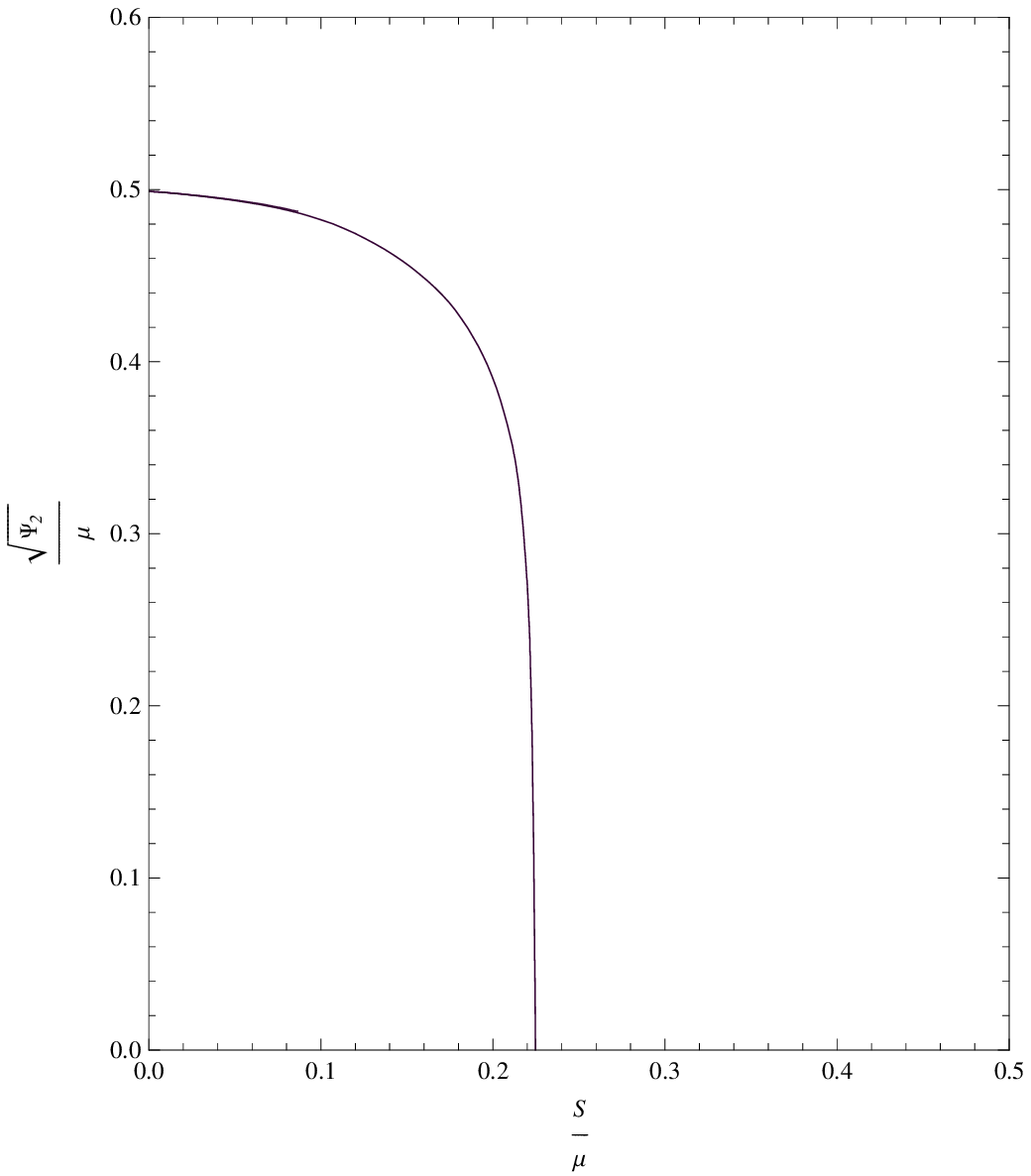}}
 \end{center}
\caption{Phase structure in presence of a non-trivial $A_x$ field. Vertical line in the left hand figure marks the first order transition.}
\label{fig:phasecurve}
\end{figure}
\subsubsection{$\mu$ fixed, $S_x$ varying}
Let us now consider the effect of turning on $A_x$. As explained in the section \ref{EABC}, turning on $A_x$ has the effect of introducing a global $U(1)$ current $J_x$ in the dual field theory. The boundary value of $A_x, S_x$ acts as chemical potential for this current. In Fig \ref{fig:phasecurve} we plot the scaled condensate strength $\sqrt{\Psi_2}/\mu$ as a function of the scaled $A_x$ chemical potential $S_x/\mu$ for different values of $1/\mu$. The plot on the left is for $1/\mu \approx 0.146$, while for the plot on the right $1/\mu \approx 0.217$. We see an interesting behaviour here: for all values of $1/\mu$ there is a critical value of the current above which there is no condensate. However, the nature of this transition seems to change with the value of $1/\mu$. For small values of this parameter, we seem to find a first order transition from the superconducting to the normal state when $S_x$ reaches a critical value $S_{x,c}$(see Fig \ref{fig:phasecurve1}). For values above a special value $1/\mu_{sp}$, the nature of the transition seems to change to second order (see Fig \ref{fig:phasecurve2}). Note that the values of $1/\mu$ or temperature that we consider are below the usual critical temperature that exists for $A_x = 0$, which is at $\frac{1}{\mu_c} \approx 0.246$ in this case.

\begin{figure}[h!]
 \begin{center}
   \subfigure[$1/\mu \approx 0.146$]{\label{fig:energycurve1a} \includegraphics[width=5.5cm]{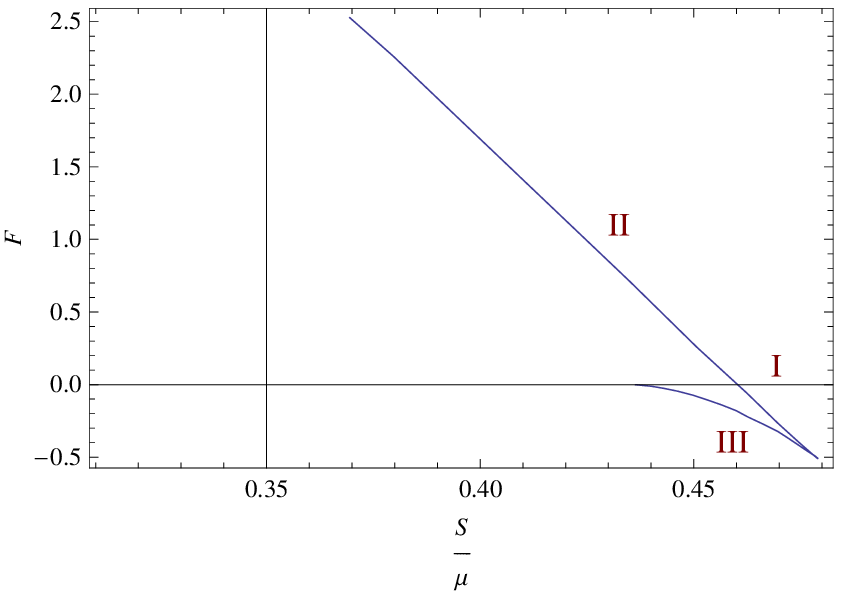}} \hspace{0.1cm}
   \subfigure[$1/\mu \approx 0.217$]{\label{fig:energycurve1b} \includegraphics[width=5.5cm]{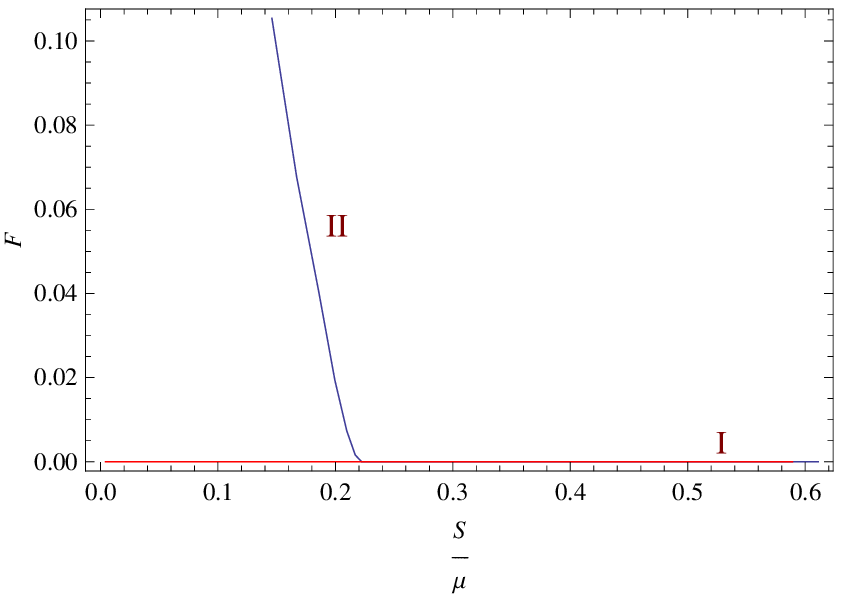}}
 \end{center}
\caption{Free energy for the different phases.}
\label{fig:energycurve1}
\end{figure}

The critical value $S_{x,c}$ can determined by comparing the free energies of the solution with supercurrent solution and the $\psi = 0$ solution for the same value of $\frac{S_x}{\mu}$. In Fig \ref{fig:energycurve1} we plot the difference in free energies of the two branches as a function of $S_x/\mu$. The figure on the left (Fig \ref{fig:energycurve1a}) is for $1/\mu = 0.146$. We see the ``swallow tail'' curve typical of first order transitions. At $S_{x,c}/\mu = 0.46$ the branches cross, and the system jumps to the normal phase where the condensate ceases to exist.  In the right hand side figure (Fig \ref{fig:energycurve1b}), we again plot the free energy difference for $1/\mu = 0.217$. We see a smooth transition to the normal phase, which is second order. The critical value of $S_x/\mu$ is $0.22$ in this case. Details of the ``swallow tail'' diagram is discussed in a similar situation in the next subsection. 

\begin{figure}[h!]
  \begin{center}
   \subfigure[$1/\mu \approx 0.146$]{\label{fig:svsja} \includegraphics[width=5.5cm]{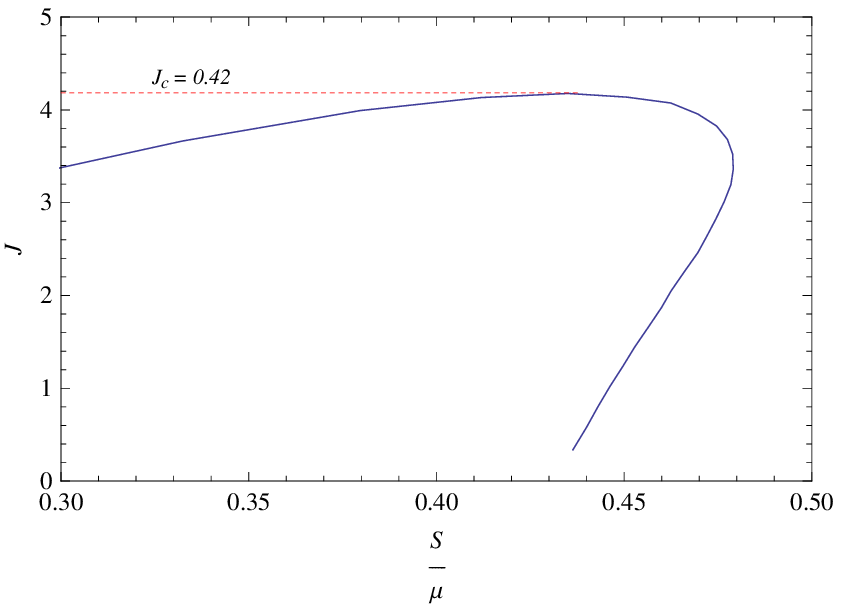}} \hspace{0.1cm}
   \subfigure[$1/\mu \approx 0.217$]{\label{fig:svsjb} \includegraphics[width=5.5cm]{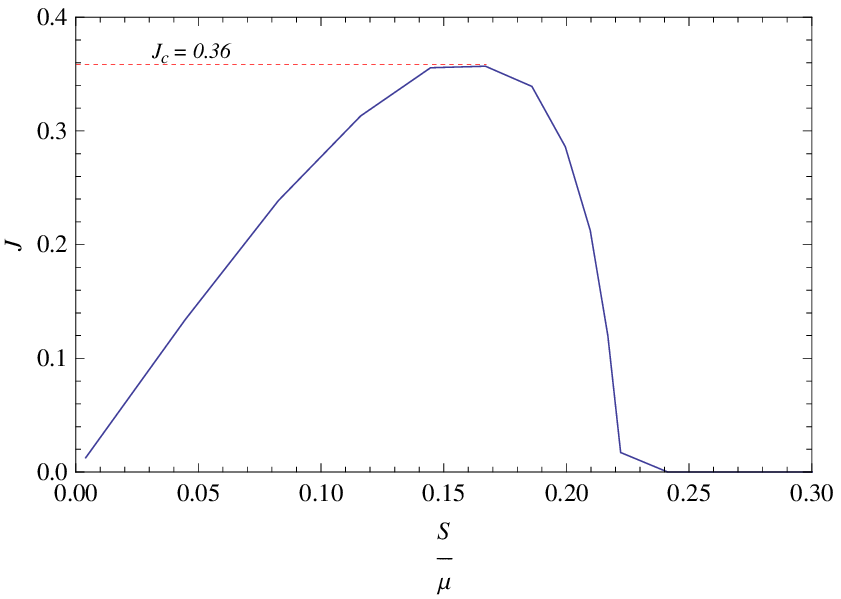}}
 \end{center}
\caption{Plot of $J_x$ as a function of $S_x/\mu$.}
\label{fig:svsj}
\end{figure}

We can also calculate the critical (or maximal) value of the ``current'' $J_x$. This can be done by reading off the current from a plot of $S_x/\mu$ vs $J_x$, shown in Fig \ref{fig:svsj}. For the two cases considered above the critical currents are $J_{x,c} = 4.2$ and $J_{x,c} = 0.36$ respectively.

\subsubsection{$S_x$ fixed, $\mu$ varying }
\begin{figure}[h!]
 \begin{center}
   \subfigure[$S_x/\mu = 0.1$]{\label{fig:fixedSa} \includegraphics[width=5.5cm]{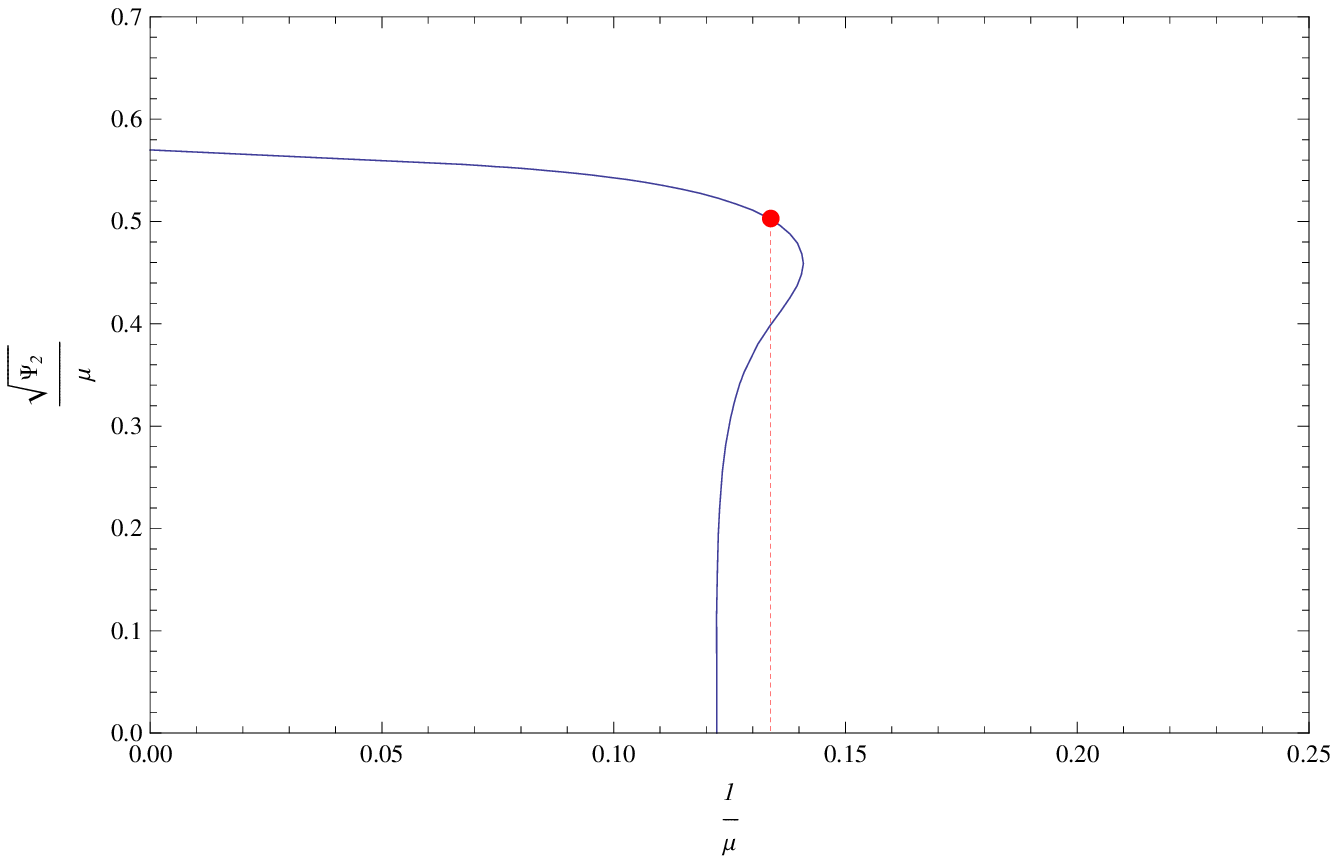}} \hspace{0.1cm}
   \subfigure[$S_x/\mu = 0.5$]{\label{fig:fixedSb} \includegraphics[width=5.5cm]{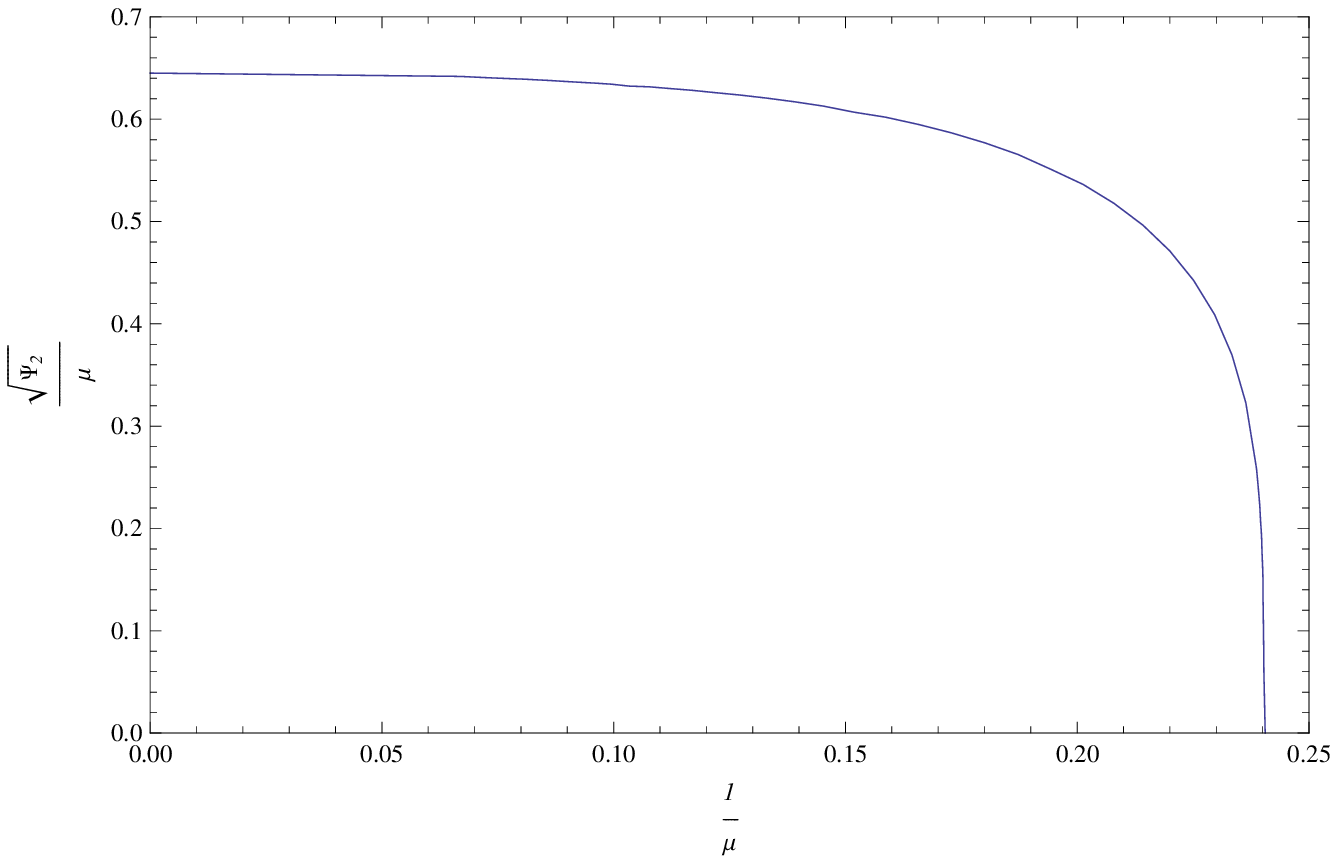}}
 \end{center}
\caption{Phase structure in presence of a non-trivial $A_x$ field. Vertical line in the left hand graph marks first order transition.}
\label{fig:fixedS}
\end{figure}

\begin{figure}[h!]
 \begin{center}
   \subfigure[$S_x/\mu = 0.5$]{\label{fig:energycurve2a} \includegraphics[width=5.5cm]{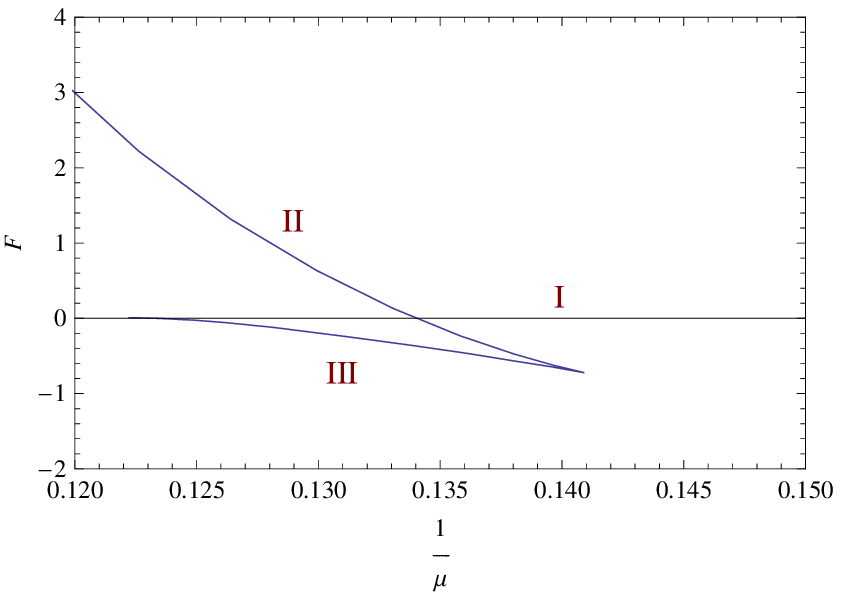}} \hspace{0.1cm}
   \subfigure[$S_x/\mu = 0.1$]{\label{fig:energycurve2b} \includegraphics[width=5.5cm]{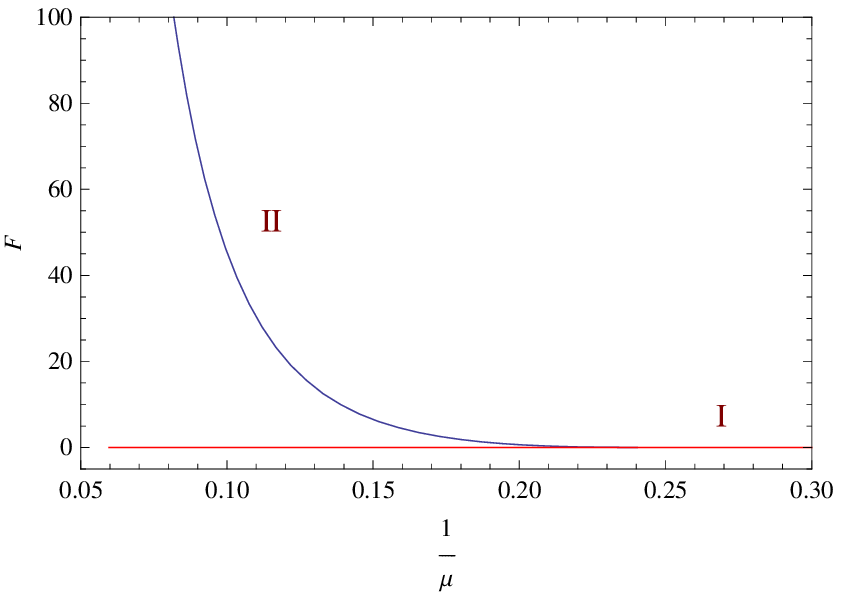}}
 \end{center}
\caption{Free energy for the different phases.}
\label{fig:energycurve2}
\end{figure}

In the above discussion, we explored the phase structure by taking constant $1/\mu$ sections. Equivalently, we can consider constant $S_x/\mu$ sections. Fig \ref{fig:fixedS} shows the variation of the condensate $\sqrt{\Psi_2}/\mu$ with $1/\mu$, with $S_x/\mu = 0.1, 0.5$ respectively for the left and right hand side plots.
The plot of the free energy difference between the normal and the supercurrent branches reveals the nature of the phase transition (see Fig \ref{fig:energycurve2}). We see again that for small values of $S_x/\mu$ the phase transition is first order ($1/\mu_c = 0.134$), and it changes to second order for higher values of $S_x/\mu$ ($1/\mu_c = 0.24$). When the transition is first order there is three branches of solution. Similar to the phase diagram of $AdS_5$ black holes in global coordinate, there is one nucleation temperature ($1/\mu_N$) where there is generation of two new solutions.  In \fref{fig:fixedSa}  $1/\mu_N \approx 0.14$. For a value of $1/\mu$ just below $1/\mu_N$, there are two possible solutions: the one with the higher value of condensate is stable (branch II), while the other one is unstable (branch III) (we have not done a local stability analysis, but this is the most likely case from the global stability.). The stable solution becomes dominant over the non-super conducting solution (branch I) at $\mu=\mu_c$. The unstable solution merges with the non-superconducting branch at an even lower value of $1/\mu$. The ``swallow tail'' diagram in \fref{fig:energycurve2a} shows this clearly. When the transition is of second order, there is no branch crossing and the non-superconducting solution becomes unstable for $\mu > \mu_c$ and the free energy of the condensate branch (II) is always less than the non-superconducting branch (I).

The critical value of the current $J_x$ can be determined similarly as above.

\begin{figure}[h!]
 \begin{center}
 \includegraphics[width=10cm]{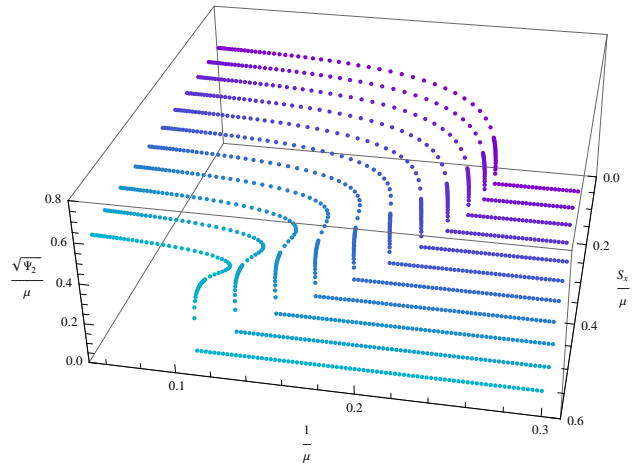}
 \end{center}
\caption{Plot of $\sqrt{\Psi_2}/\mu$ as a function of $1/\mu$ and $S_x/\mu$.}
\label{fig:3dcurve}
\end{figure}

In order to better understand the phase structure we can look at a three dimensional plot showing the variation of the three relevant dimensionless parameters in this case namely $1/\mu, S_x/\mu$ and $\sqrt{\Psi_2}/\mu$. This is shown in Fig \ref{fig:3dcurve}. The curves in Fig \ref{fig:phasecurve} are constant $1/\mu$ sections of this plot, while Fig \ref{fig:fixedS} shows constant $S_x/\mu$ sections. We can clearly see the change in the behaviour of $\sqrt{\Psi_2}/\mu$ as a function of $S_x/\mu$ with changing $1/\mu$. Below $1/\mu_{c'} = 0.22$ the dependence is non-monotonic, indicative of a first order transition. Above this value the dependence becomes monotonic and we have a second order phase transition.
\subsubsection{Phase boundary}
We can also look at the phase structure on the $S/\mu$, $1/\mu$ plane. The following figure (Fig \ref{fig:phaseplot}) shows the result. The blue line indicates the region of second order phase transitions (the line in the right hand sign of the dot), which changes into first order at the red line (the line at the left hand side of the dot). The area enclosed by the transition line and the axes represent the condensate phase, while above the line the system is in the normal phase. The intercepts of this curve with the axes define two critical points: at $S_x = 0$, there is a second order phase transition as $1/\mu$ is increased at $1/\mu_c = 0.246$, while near $1/\mu = 0$ there is a first order phase transition with increase in $S_x/\mu$ at $S_{x,c}/\mu = 0.874$.
\begin{figure}[h!]
 \begin{center}
 \includegraphics[width=10cm]{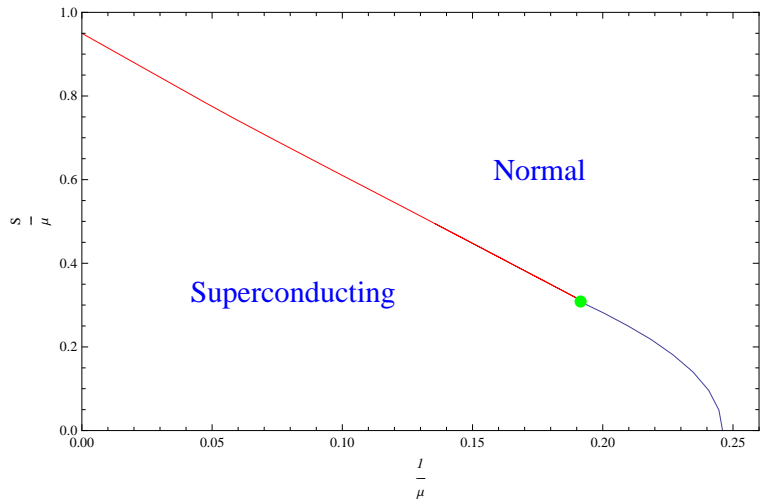}
 \end{center}
\caption{Phases of the Abelian Higgs model. The nature of the phase transition changes from second order (blue line) to first order (red line) at the ``special point'' (green dot).}
\label{fig:phaseplot}
\end{figure}

\subsection{$\Psi_2 = 0$}
\begin{figure}
  \begin{center}
  \includegraphics[width=8cm]{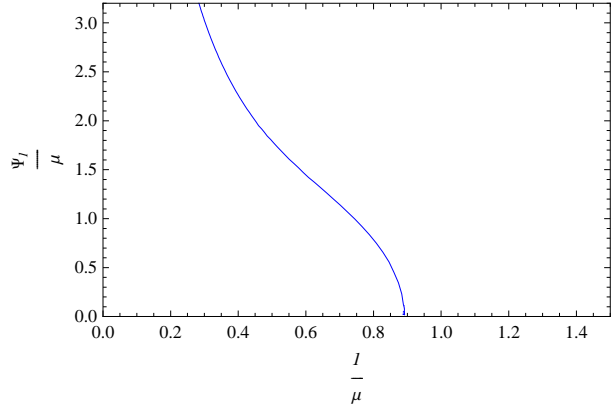}
  \end{center}
\caption{Plot of $\Psi_1$ as a function of $1/\mu$, for $A_x = 0$.}
\label{fig:psi1curve}
\end{figure}

We now consider the case where $\Psi_2 = 0$. This corresponds to the boundary condition when $\psi(z) \sim z$ as $z \rightarrow 0$. Then for $A_x = 0$ we get the dependence of the scaled condensate strength $\Psi_1/\mu$ on $1/\mu$ shown in Fig \ref{fig:psi1curve}.
Near $1/\mu = 0$ the condensate strength diverges\footnote{Note that for large values of the condensate $\Psi_1$ the gravity backreaction is important and our approximation is no longer valid.}, while near the critical value of $1/\mu = 1/\mu_c$ the curve has the dependence $\Psi_1/\mu \sim (1/\mu - 1/\mu_c)^{1/2}$ as before. So we again have a second order phase transition at this point. The critical value $1/\mu_c = 0.89$.

\begin{figure}[h!]
 \begin{center}
   \subfigure[$1/\mu = 0.61$]{\label{fig:phasecurve1_psi1} \includegraphics[width=5.5cm]{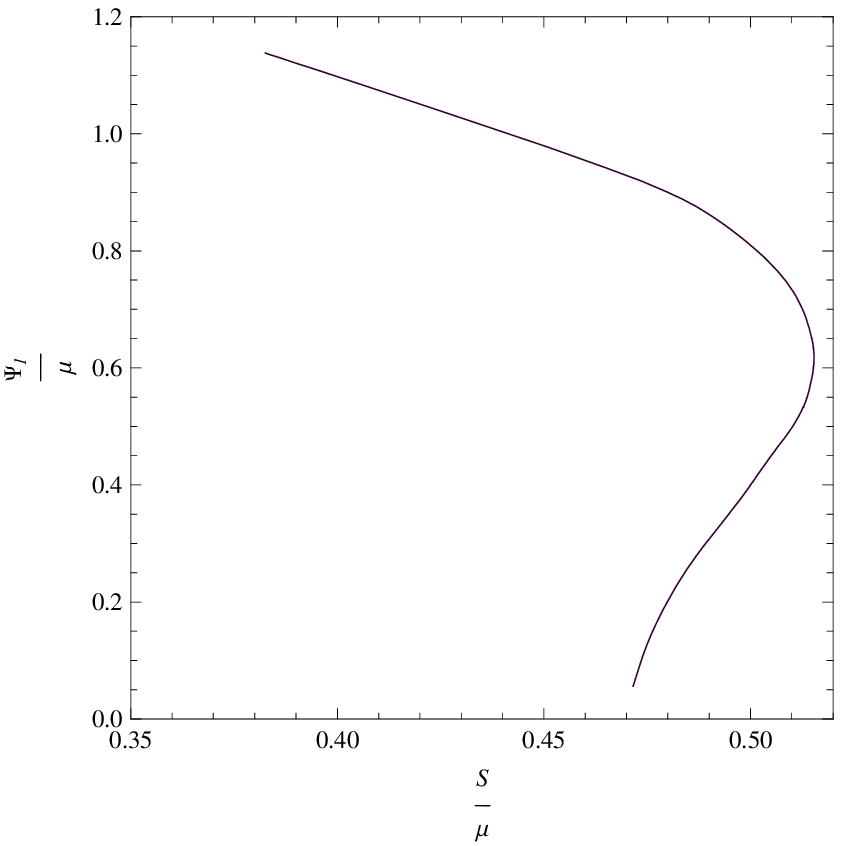}} \hspace{0.1cm}
   \subfigure[$1/\mu = 0.813$]{\label{fig:phasecurve2_psi1} \includegraphics[width=5.5cm]{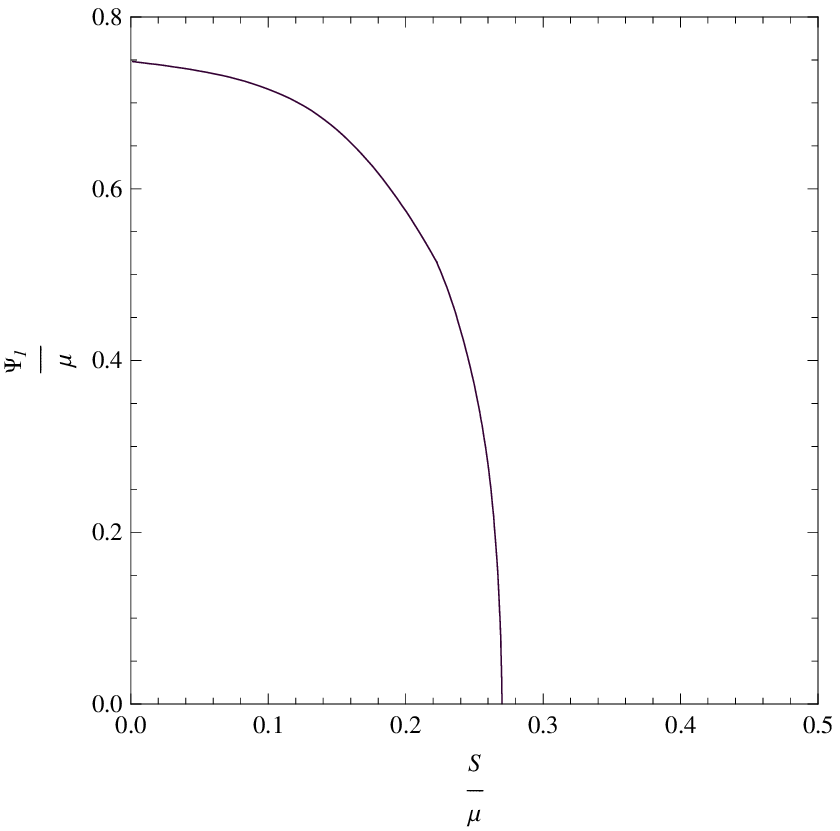}}
 \end{center}
\caption{Plot of $\Psi_1$ as a function of $1/\mu$, for $A_x \neq 0$.}
\label{fig:phasecurve_psi1}
\end{figure}
We can now turn on $A_x$ in this setting, and investigate the response of the system as we increase the chemical potential for the $A_x$ field, $S_x$ at various values of $1/\mu$. The results are shown in Fig \ref{fig:phasecurve_psi1}. The story here is qualitatively similar to the $\Psi_1 = 0$ case: we again see the existence of a critical current above which there is no condensate. For small values of $1/\mu$ there is a first order phase transition as $S_x$ is varied, as seen on the left hand side plot (Fig \ref{fig:phasecurve1_psi1}). Here $1/\mu = 0.61$. For larger values of $1/\mu$ the nature of the transition changes to second order, as seen on the right hand side plot for $1/\mu = 0.813$ (Fig \ref{fig:phasecurve2_psi1}). Note again that we are below the critical value $1/\mu_c = 0.89$ where the condensate ceases to exist for $A_x = 0$.

\begin{figure}[h!]
 \begin{center}
 \includegraphics[width=10cm]{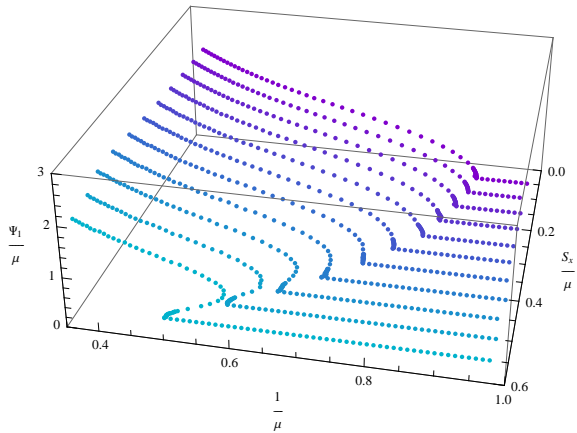}
 \end{center}
\caption{Plot of $\Psi_1/\mu$ as a function of $1/\mu$ and $S_x/\mu$.}
\label{fig:3dcurve_psi1}
\end{figure}

As in the $\Psi_1 = 0$ case, we can look at the behaviour of $\Psi_1/\mu$ as a function of $1/\mu$ and $S_x/\mu$, shown in Fig \ref{fig:3dcurve_psi1}. Below $1/\mu_{c'} \approx 0.6$, $\Psi_1/\mu$ is a non-monotonic function of $S_x/\mu$ and we have a first order transition at a critical value $S_{x.c}/\mu$. For $1/\mu > 1/\mu_{c'}$ the dependence becomes monotonic, and the transition becomes second order.

\section{Connection to Superfluids}
\label{PHYREL}
We note here that due to the global nature of the associated $U(1)$ symmetry, the condensation phenomenon we witness here is more similar to a superfluid. The phase structure we find can be interpreted by comparing with the known behaviour of superfluids \cite{Tinkham}. Consider a charged superfluid confined to a thin film. The system can be described by a Landau-Ginzburg type theory with the following Lagrangian:
\begin{equation}
 L = \alpha(T)|\psi|^2 + \frac{\beta}2 |\psi|^4 + \frac{1}{2m} \left| \left(-i\nabla - \mathbf{A}\right) \psi \right|^2,
\end{equation}
where $\psi$ is the condensate. The supercurrent can be defined in terms of $\psi(\mathbf{r}) = |\psi(\mathbf{r})| e^{i\varphi(\mathbf{r})}$ as
\begin{equation}
 \mathbf{J} = \frac{|\psi|^2}{m} \left(\nabla \varphi - \mathbf{A} \right) \equiv |\psi|^2 \mathbf{v}_s,
\end{equation}
where $\mathbf{v}_s$ is the mechanical fluid velocity. We can think of $\mathbf{J}$ as being the response to the mechanical velocity $\mathbf{v}_s$. Note that in the field theory we get from our gravity picture, both the condensate $\psi$ and the vector potential $\mathbf{A}$ are spatially homogeneous. Then $\mathbf{v}_s \propto \mathbf{A}$, i.e., a constant mechanical velocity is induced by a uniform vector potential\footnote{Note that since our system is homogeneous, we can make the direction along the vector potential periodic without any modifications. A constant vector potential cannot be gauged away and is thus physically meaningful. We thank Gary Horowitz for discussions on this point.}. 

At low temperatures the quasiparticle energies are shifted by an amount proportional to the fluid velocity $\mathbf{v}_s$, and this reduces the energy gap as $\mathbf{v}_s$ is increased. There exists a critical velocity where the gap goes to zero; at this point the system undergoes a first order transition to the normal state. Experimentally, it is known for superfluids at temperatures close to zero that as $\mathbf{v}_s$ is increased, the supercurrent $\mathbf{J}$ initially increases in proportion to $\mathbf{v}_s$. However, once $\mathbf{v}_s$ reaches a critical value $\mathbf{v}_{s,c}$, the current drops steeply to zero. This corresponds to the first order phase transition at the critical velocity.

The situation is quite different near the critical temperature. There is still a phase transition, but it can be shown to be of second order. The current drops to zero smoothly at the critical velocity. 

These observations agree qualitatively with our model if we identify the quantity $S_x$ with the magnitude mechanical velocity of the fluid $\mathbf{v}_s$. 

\section{Gravity backreaction}
\label{BCKR}
It is an important question that how the solutions change as we incorporate the gravity backreaction. Without the scalar condensation such a solution is just the standard RN black hole in $AdS^{3+1}$ space. The metric of which is given by \cite{Romans:1991nq},
\bea
ds^2 &=& \frac{L^2}{z^2} (-f(z) dt^2+dx^2+dy^2)+\frac{L^2 dz^2}{z^2 f(z)}   \\
\nn A_t &=& \mu(1-\frac{z}{z_0}) \\
\nn f(z) &=& 1+ q^2 z^4-(1+q^2) z^3 
\eea 
The whole solution including the $A_t$ can be given a Lorentz boost with velocity $v$ in the $x$ direction and the resulting solution has a $A_x^{new}$ given by $A_x^{new}=\beta A_t$ , where $\beta=\frac{v}{\sqrt{1-v^2}}$. As $A_t$ has a non-trivial dependence on the radial coordinate $z$, $A_x$ will also have the same non-trivial dependence. Dual of such a configuration is naturally interpreted as a boosted gauge theory plasma with charge. This solution may also be thought as the ``backreacted'' version of our constant $A_x$ solution in a normal (non-superconducting) black hole background. 
\par
Now, let us assume the likely scenario that a superconducting black hole solution survives after considering gravity backreaction. Such a backreacted superconducting solution may also be given a Lorentz boost and the resulting solution will have a non-trivial dependence on $A_x$. Our supercurrent solution should \emph{not} be confused with such a trivial boosted solution. Our  solution should be interpreted as a solution where black hole horizon remains fixed (or is moving with a constant velocity) but the condensate has an arbitrary velocity with respect to the horizon. The supercurrent solution can not be generated by boosting, as it does not obey the constraint of a boosted solution, i.e. $A_x^{new}=\beta A_t$ and $g_{tx}^{new}=-\beta ( g_{tt}+g_{xx})$. With a chemical potential $S_x$, it is likely that the dominant solution at low temperature will be a supercurrent type solution. Whether the structure of the phase diagram changes significantly after considering gravity backreaction is an open question.

\section{Conclusions}
\label{CON}
In this paper we exhibit a static solution to the system with a charged scalar field coupled to the AdS black hole, which in the dual field theory corresponds to a static current flowing in a superconducting fluid with no emf applied. We see an interesting phase structure, with a first order transition to the normal state at low temperatures as the fluid velocity is increased. At temperatures close to the critical value $T_c$, the transition becomes second order. As mentioned in the previous section, it would be nice to verify whether the phase diagram is modified in the fully back reacted geometry.

The model we have considered does not have any magnetic field in the boundary, as $A_x$ does not depend on any of the field theory directions $y$. One can turn on a non-trivial magnetic field by incorporating a dependence on $y$ in $A_x$, i.e., $A_x \equiv A_x(z,y)$. This can be used to study phenomena such as the Meissner effect. In particular, one can check the nature of the superconductor, i.e., whether it is Type I or Type II. In the latter case one can try to find vortex solutions. However, in this case the field equations become coupled nonlinear partial differential equations, which are harder to solve. It would also be interesting to study the modifications of these models to include impurity etc. 

Embedding the superconducting gravity solutions into string / M theory is an important issue. Whether any probe brane configuration in some AdS like space gives rise to the type of Lagrangian we are discussing, would be an interesting avenue to explore. 

As we have discussed, the $U(1)$ symmetry in our model is a realized globally in the boundary. It will be an interesting direction to setup some brane/gravity model where the symmetry breaking is local. That would be more akin to real life superconducting materials.  

We have just begun to understand strongly coupled physics of condensed matter systems holographically. Exploring these modifications may allow us to extract new information about the universality classes and phase structures of strongly coupled systems.

\section{Acknowledgements}
We would like to thank Mark van Raamsdonk for helpful discussions and comments on our work. We would also like to thank Ian Affleck, Adrian Giuseppe Del Maestro, Marcel Franz, Eran Sela, Ariel Zhitnitsky for fruitful conversations on aspects of condensed matter physics. We thank Gary Horowitz, Spenta Wadia, Wen-Yu Wen and Henry Ling for discussions and comments. We thank the String Theory Group at UBC for their support and encouragement, and the organizers of the PIMS string conference at BIRS, Banff. PB and AM acknowledge support from the Natural Sciences and Engineering Research Council of Canada. HHS is supported by the UBC University Graduate Fellowship.

\appendix
\section{A note on dimensions}
Lets start with the metric
\begin{equation}
 ds^2 = -f(r) dt^2 + \frac{dr^2}{f(r)} + r^2 (dx^2 + dy^2)
\end{equation}
Here $ds^2$ is dimensionless, hence if we scale the boundary coordinates $x, y, t$ by a constant $\alpha$, then $r$ needs to be scaled by $\alpha^{-1}$. Thus $f(r)$ should scale as $\alpha^{-2}$. From the equation for $A_t$
\begin{equation}
 A_t'' + \frac2r A_t' - \frac{2\Psi^2}{f}A_t = 0
\end{equation}
we see that $\psi$ does not need to scaled. Also, from the equation for $\psi$
\begin{equation}
 \psi'' + \left( \frac{f'}{f} + \frac2r\right) + \frac{A_t^2}{f^2}\psi + \frac{2}{L^2f}\psi = 0
\end{equation}
we see that $\psi''$ and $\frac{A_t^2}{f^2}\psi$ must scale the same way, so $A_t$ scales like $r$ (since $f(r)$ scales like $r^2$). Now we know the boundary coordinates $x, y, t$ have mass dimension $-1$. From the scaling behaviour, we can now determine the dimensions of all other operators:
\begin{eqnarray}
 \nonumber \left[\psi\right] &=& 0 \\
 \left[r\right] = \left[A\right] = \left[\mu\right] = \left[\Psi_1\right] &=& 1 \\
 \nonumber \left[\rho\right] = \left[\Psi_2\right] &=& 2
\end{eqnarray}

\bibliographystyle{hunsrt}
\bibliography{sucu.bib}
\end{document}